\documentclass[conference]{IEEEtran}
\usepackage{cite}
\usepackage{amsmath,amssymb,amsfonts}
\usepackage{algorithmic}
\usepackage{graphicx}
\usepackage{textcomp}
\usepackage{xcolor}

\usepackage[utf8]{inputenc}

\begin{document}

\title{Secondary Inputs for Measuring User Engagement in Immersive VR Education Environments}
\author{\IEEEauthorblockN{1\textsuperscript{st} David Murphy}
   \IEEEauthorblockA{\textit{School of Computer Science and Information Technology} \\
   \textit{University College Cork}\\
Cork, Ireland \\
d.murphy@cs.ucc.ie}\\
\and
\IEEEauthorblockN{2\textsuperscript{nd} Conor Higgins}
\IEEEauthorblockA{\textit{School of Computer Science and Information Technology} \\
\textit{University College Cork}\\
Cork, Ireland \\
}
}

\maketitle

\begin{abstract}

This paper presents an experiment to assess the feasibility of using secondary input data as a method of determining user engagement in immersive virtual reality (VR). The work investigates whether secondary data (biosignals) acquired from users are useful as a method of detecting levels of concentration, stress, relaxation etc. in immersive environments, and if they could be used to create an affective feedback loop in immersive VR environments, including educational contexts. A VR Experience was developed in the Unity game engine, with three different levels, each designed to expose the user in one of three different states (relaxation, concentration, stress). While in the VR Experience users' physiological responses were measured using ECG and EEG sensors. After the experience users completed questionnaires to establish their perceived state during the levels, and to established the usability of the system. Next a comparison between the reported levels of emotion and the measured signals is presented, which show a strong correspondence between the two measures indicating that biosignals are a useful indicator of emotional state while in VR. Finally we make some recommendations on the practicalities of using biosensors, and design considerations for their incorporation in to a VR system, with particular focus on their integration in to task-based training and educational virtual environments.
\end{abstract}


\section{Introduction}
Virtual Reality is an immersive medium where the goal is for the user to believe that they are in a different reality. Conventionally we use various instruments, such as surveys and questionnaires, to measure the level of immersion, presence and the emotional state of users in VR~\cite{Schuemie:2001wb,Usoh:2000bp,Witmer:1998bn,Insko:2003us,Pausch:1997:QIV:258734.258744,Schwind:2019fq}. With recent developments in physiological sensor technology ~\cite{Wiederhold:2001uq} we maintain that the incorporation of physiological signals (biosignals) can provide an objective measure of a user's performance and emotional state while in VR and serious games~\cite{sliney2009secondary,crowley2010evaluating,sliney2009assessment}.

The current experiment examines the use of users' `secondary input', which can be fed back into an immersive experience, to create an affective and reactive feedback loop, or can be used as part of a summative assessment of a student's participation in a VR Education Experience.

To begin we should define what we mean by the different types of assessment/data inputs. If a user is playing a game or participating in an immersive experience with a controller, they are consciously, directly controlling it  -- this is what we term as a `primary input'~\cite{sliney2009secondary}. 
Secondary input can be thought of as `passive input' or latent signals that a user generates while interacting in VR or an immersive virtual education environment. These secondary inputs can be divided into two categories, (a) secondary (performance) data sources derived from interaction, such as time-on-task and number-of-errors, which are a measure of how well a user is performing cognitively with a task or environment, and (b) endogenous physiological signals captured from the user, which can be used to objectively measure how a user is engaging with the task or environment~\cite{crowley2010evaluating, sliney2009assessment}.    
Endogenous secondary input can be thought of as physiological responses and reactions to stimuli within the VR experience. 

The current work focuses on the endogenous secondary inputs from the physiological sensors, investigating the use of biosignals while a user is participating in an immersive VR experience. We analyse the signals to determine whether or not they are robust enough to be used as an input source to measure the user's engagement with the VR environment and to enhance the learning experience.
\section{Literature Review}

As the final outcome would be to use secondary input signals in order to enhance immersion, it is important to understand first the current level of immersion in modern VR experiences, and why it is important. 

The use of immersion in video and serious games often enhances the play experience, especially in games that require a certain level of suspension of disbelief, such as fantasy or science fiction games. Author and game designer Ernest Adams defines immersion as one of three distinct kinds; tactical immersion, strategic immersion, and narrative immersion~\cite{adams2004postmodernism}. Adams defines tactical immersion as immersion found in the moment-to-moment gameplay, the feeling of being ``in the zone'', after completing tasks in game, achieved by slick, responsive user interfaces. 
Strategic immersion is associated with mental challenge, by coming up with strategies to solve problems, and optimising solutions to achieve victory, something that is very relevant in the development of serious games and educational VR experiences. Finally, narrative immersion is described as being invested in a story and in interesting characters, similar to a book or a film. 

In addition to these three forms of immersion, a fourth was identified by Bj\"{o}rk and Holopainen, which is more relevant to VR than the others; spatial immersion~\cite{holopainen2005patterns}. Spatial immersion is recognised as feeling as though the user is physically present in the scene, feeling that the experience is realistic. Each of these types of immersion are important when developing a game or experience, but spatial immersion can be crucial for developing VR; if a virtual environment is not immersive enough, or not perceptually convincing enough for users, it may interfere with the learning, as users may feel they are taken out of the experience. While these levels of immersion are important in helping an experience feel more realistic, they are often challenging to create. Before it was more widely and commercially available, there was some scepticism about the ability of VR to provide a decent level of immersion. Early experiments showed that using VR devices to perform cognitive tasks (in one particular case, searching for a specific target inside a room) did not necessarily provide a substantial benefit over using a standard desktop~\cite{Pausch:1997:QIV:258734.258744}. However, if was found that if the target was not present in the scene, then VR users were quicker in asserting that fact, suggesting that using the VR system helped users built a better spatial frame of reference that non-VR users. Expanding on this study, Pausch et al. explored the same findings, while attempting to expand the type of testing, and determine whether different types of VR, e.g. Desktop VR and Head-Mounted Display (HMD) VR, or different forms of cognitive trials could affect a user's level of immersion~\cite{Pausch:1997:QIV:258734.258744}. As the technology behind VR has evolved, along with the evolution of storytelling and visual graphics in games, more and more studies have been undertaken in determining just how immersive VR can, or should be.

\section{Biosignals}
As mentioned earlier, we believe that the incorporation of biosignals as sources of secondary input data provides objective measures of a user's physiological response to events and stimuli in a virtual environment. In certain types of VR education experiences it is important to know whether the user is stressed or not, or whether the user is concentrating adequately, which are of particular interest in domains such as medical and health education, aviation training, or task-based training systems.
\subsection{EEG}
The Electroencephalogram (EEG) signal is a representation of the brain's electrical activity -- the electrical impulses of the brain's cells communicating with one another. The EEG has a number of medical uses, such as checking for conditions such as epilepsy, studying sleep disorders, and detecting general brain activity. An EEG signal is typically measured using electrodes that are placed on the scalp. The brainwave output gathered from an EEG is actually a complex compound signal that needs to be broken down into a number of different frequency bands, each with a corresponding wave (frequency bandwidth) which is associated with a different mental or cognitive function. These waves are delta, theta, alpha, beta and gamma waves~\cite{MARCUSE201639}.  Table ~\ref{freqBands_table} shows each wave's frequency band, as well as some of the states it is generally associated with.

\begin{table}[htp]
\centering
\begin{tabular}{|l|l|l|}
\hline
Wave & Frequency (Hz) & States \\
\hline
Delta ($\delta$) & $0.1 - 4$ & Deep sleep \\
Theta ($\theta$)& $4 - 8$ & Daydream, trance, drowsiness \\
Alpha ($\alpha$)& $8 - 13$ & Conscious relaxation, meditation \\
Beta ($\beta$)& $13 -30$ & Focused, engaged, attention, thinking \\
Gamma ($\gamma$)& $>30$ & Cognition, perception, memory\\
\hline
\end{tabular} \caption{The different brainwaves, frequency bands, and associated states.}
\label{freqBands_table}
\end{table}

In recent years, there have been a number of papers published exploring the usefulness of these brainwaves, in both detecting certain physiological states of a user and of attempting to influence a virtual experience. The studies undertaken investigated the use of commercial, entertainment focused EEG devices, rather than medical-grade probes, as these would be too expensive for regular users. Testing on users showed that while they were not up to standard for use as a medical device, commercial grade devices such as those created by NeuroSky and Emotiv could be useful in BCI (brain-computer interface) applications~\cite{Wojciech:2014}. Subjects having their EEG signal monitored showed results that were indicative of the expected response for each brainwave, e.g. when placed in a relaxing environment, users' alpha waves were reported to have higher activity. These devices were tested under a number of different conditions, to assess not only their accuracy in retrieving the EEG data, but also in their reliability; if they data could be recaptured in a number of different situations, it would be very possible that it could be used in such applications to measure a user's cognitive state. Results showed that devices like these could be highly useful in gathering individual EEG readings, for applications other than medicine or health~\cite{ROGERS201687}. 

\subsection{ECG}
An Electrocardiogram (ECG) records the electrical signals of the heart. Each beat of the heart is triggered by an electrical impulse normally generated in the upper right chamber of the heart. Sensors (electrodes) are placed on the body to detect this electrical activity.  An ECG records the timing and strength of these signals as they travel through the heart. From these recordings we can determine a number of characteristics of the performance of the heart, including the heart rate and rhythm, and its variability (HRV), which are indicators of stress~\cite{Kim:2018go} and mental workload~\cite{Meshkati:1988ev}. ECGs, as well as other forms of heart performance measurement, have become more common in immersive virtual environments~\cite{Malinska:2015bt}. In the current work, the inclusion of an ECG signal would help in determining whether or not the EEG signal would be a viable and robust source of secondary input.

The sensors for each signal were chosen based on their ergonomics, ease of use, and the reliability of the signal received from them. For the EEG signal, the chosen sensor was the MyndPlay BrainBand, a consumer and research grade EEG; it outputs the wearer's raw brainwave data (the EEG signal itself) at a sampling frequency of 512Hz. The device consists of a single electrode placed on the user's left temple, along with an accompanying ear clip which was affixed in order to cancel out the electrical noise produced by the body. This probe was attached via a Velcro strap to the rim of a VR headset (Oculus Rift) in this project.
For the ECG we used AliveCor's Kardia, a device that can provide medical-grade ECG readings, with a very small and sleek form factor, about the thickness of a coin and the size of a large plaster. To use it, the user  places one to two fingers on one of two metallic pads (one per pad per hand), which then allows an ECG to be read. Designed to work as a highly mobile device, it connects to a Kardia Mobile app for iOS and Android, which displays the user's ECG signal in real time. The form factor, ease of use, and medical accuracy of the device meant that it was a very good fit for the project, and thus it was decided that it would be used as the second source of secondary data to accompany the BrainBand's EEG signal.

\section{Methods}
A three-scene Virtual Reality environment was constructed in Unity 3D that incorporated the MyndPlay BrainBand EEG and the AliveCor Kardia ECG.  
The three-scenes represented relaxation, concentration, and stress respectively. The experience also included an introductory level, where users are introduced to VR and the navigation scheme, which afforded an opportunity to get a base-level reading for the EEG and ECG sensors while the user is in a neutral scene.

\subsection{Scene 1 - Relaxation}
A simple free exploration Forest-based scene was designed to make the user feel relaxed and at ease. The scene included ambient outdoor music and was accompanied by gradual transitions and a pleasant voiceover.  A number of camera image effects were added, in order to soften the lighting, correct the colour of the scene by making it warmer, and add a slight blur, in addition to adding some sunbeams.

\subsection{Scene 2 - Concentration}
\begin{figure}[htbp]
\centering
\includegraphics[width=\linewidth]{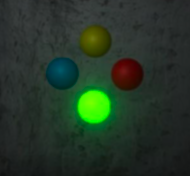}
\caption{The Simon Says glowing orbs which the user must reproduce on the XBox Controller}
\label{SimpleSimonSays}
\end{figure}

The second scene was designed to have the user focus on a specific mental task, in order to observe if a level of concentration could be determined.
This was achieved using a “Simon Says” style game, in which the user is presented with a specific sequence of glowing objects (orbs), and was asked to reproduce the pattern by pressing the corresponding coloured buttons on the Xbox controller. As the pattern went on, the rate of the sequence increases, thus requiring the users to concentrate more and focus on getting the button sequence correct (see figure~\ref{SimpleSimonSays}).

\subsection{Scene 3 - Stress}

The third and final scene was designed to place the users in an environment where they would feel stress and tension. The scene was modelled on the inside of an `abandoned'  spaceship, with dark metallic interiors, narrow corridors, flashing lights, and foreboding spatialised sound effects. Other cinematic devices such as camera effects, mechanical noises, and so on were used to heighten tension.

\subsection{Integration of Devices}
The BrainBand connected to the system via software called MyndPlayer over Bluetooth. The software provides access to the raw EEG data at 512Hz and proprietary ``eSense'' data, which measures of attention, focus and mediation as determined by the Neurosky algorithm. In this research we used the raw EEG, which was subsequently processed and split in to frequency bands using an FFT in MATLAB.

\begin{figure}[htbp]
\centering
\includegraphics[width=\linewidth]{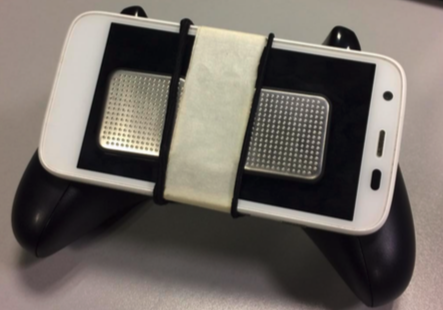}
\caption{Kardia ECG attached to Android Phone attached to XBox Controller}
\label{phoneKardia}
\end{figure}

The Kardia device transmits the ECG data as an ultrasonic signal which is picked up by a phone’s microphone and processed within the mobile app; it sends a frequency modulated (FM) signal containing the ECG data~\cite{albert2012wireless}. From the patent, it was found that the device was sending the data using a 19kHz carrier wave, in order to keep the signal immune to ambient noise. This carrier wave was then modulated with the 200Hz/mV ECG signal, with a range of ±5mV, giving a signal of between 18 and 20kHz. The signal was run through a sound check application to confirm that it was indeed transmitting a signal at 19kHz. When integrating the Kardia in to our setup, we had to attach it to a controller that ensured both hands were in contact at all times, hence we used the XBox controller that shipped with the Oculus Rift. Next we used an Android phone to receive the FM audio signal, demodulate and  send it over Bluetooth to our VR workstation. Fig~\ref{phoneKardia} illustrates the arrangement.

\subsection{Participants}
Twenty one subjects (n=21, 16 Male, 5 Female) from a population of University students.  Table~\ref{vrExperience} shows VR Experience as a percentage of subjects. None of the participants reported having a pre-existing heart condition (e.g. arrhythmia, tachycardia).

\begin{table}[htp]
\centering
\begin{tabular}{|l|l|}
\hline
Amount & Percentage \\
\hline
None & 23.8\%\\
Low & 38.1\% \\
Some & 14.3\% \\
Much & 23.8\%\\
\hline
\end{tabular} 
\caption{VR Experience.}
\label{vrExperience}
\end{table}

\subsection{Procedure}
Subjects participated individually in the experiment. Subjects were told that the purpose of the experiment was to determine the ergonomics of using the physiological sensors while in VR. The researcher explained the devices as they were attached to the subject. When the subject entered in to the Virtual Environment they first went through an Introductory Level, which conditioned them to this particular implementation of VR and associated control scheme, and also served to act as a ground-truth for the sensors. Next the subject undertook the three-scene experience. Following the experiment subjects completed a post-test questionnaire and a System Usability Survey.
\section{Evaluation and Results}
Early analysis of the results from the physiological sensors are promising, however there are some caveats with this data. When attaching a single-electrode EEG on to a Head Mounted Display (Oculus Rift) one has to ensure that the arrangement is comfortable, while at the same time providing a stable/reliable signal reading. As we subsequently found out, excessive movement by the user, and of the user's head introduced noise in to the raw signal. With respect to the ECG, to maintain the electrical circuit it is paramount that the user keeps both hands in direct contact with the respective sensor. During the experiment there were occasions where that contact was momentarily broken which led to a loss of data. 
\subsection{ECG}
\begin{figure}[htbp]
\centering
\includegraphics[width=\linewidth]{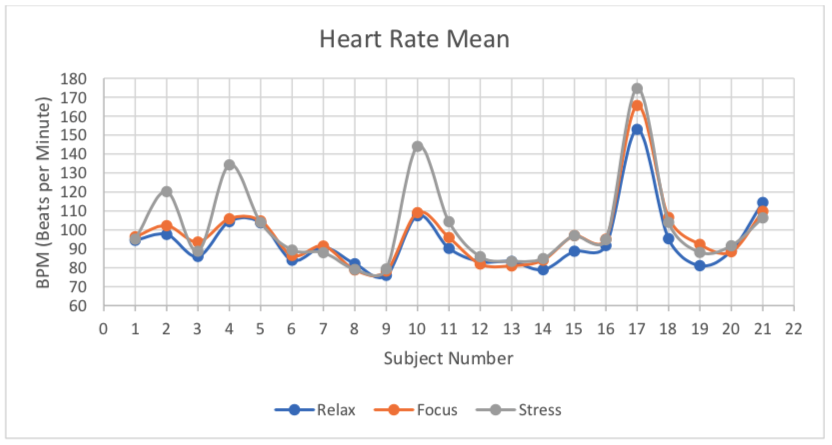}
\caption{Heart Rate mean values for the three scenes.}
\label{HeartRateMean}
\end{figure}

The heart rate data shows strong indications of an increased heart rate in the stressful scene. Interestingly, there was also indications of an elevated heart rate during the concentration/focus scene; this may be as a result of the mental workload involved in the scene, or associated with frustration each time the user was beaten in the memory task.
Figure~\ref{HeartRateMean} illustrates the mean heart rate per subject, over the three different scenes (Relaxed, Focused/Concentration, and Stressed).

\subsection{EEG}
\begin{figure}[htbp]
\centering
\includegraphics[width=\linewidth]{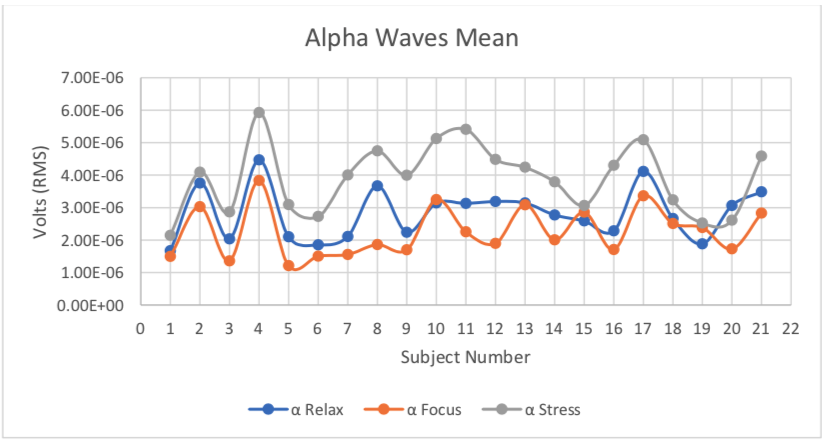}
\caption{Alpha wave mean values for each subject.}
\label{AlphaWavesMean}
\end{figure}

The results from the EEG signal were slightly more ambiguous. The main elements being looked for were: an elevated alpha level for relaxation, an elevated beta level for concentration/focus, and an elevated theta level and reduced alpha level for stress. In Figure~\ref{AlphaWavesMean}, while the alpha waves did have a larger magnitude in the relaxation level, they did not appear to be reduced in the stress level; this may have been due to the fact that the participants tended to move their heads a lot more in the stress level, as they were constantly trying to examine their surroundings. However, most participants had a larger alpha wave in the relaxation scene over the concentration scene, both of which had far less head movement. This, combined with the heart rate and answers from the questionnaire may point towards the alpha wave being a good indicator of relaxation.

\begin{figure}[htbp]
\centering
\includegraphics[width=\linewidth]{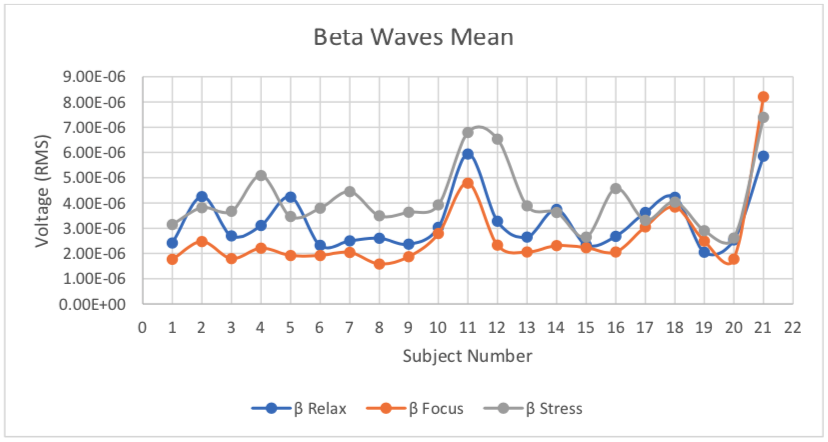}
\caption{Beta wave mean values for each subject.}
\label{BetaWavesMean}
\end{figure}

In terms of concentration, it was slightly more difficult to determine. While generally it is asserted that the beta wave is associated with concentration~\cite{seo2010stress}, the EEG data received did not show a marked increase in beta wave activity in the concentration scene over the other two scenes (figure~\ref{BetaWavesMean}). Perhaps this was once again due to an increase in head movement; the concentration scene seemed to be the scene in which the users were the most stationary.

\begin{figure}[htbp]
\centering
\includegraphics[width=\linewidth]{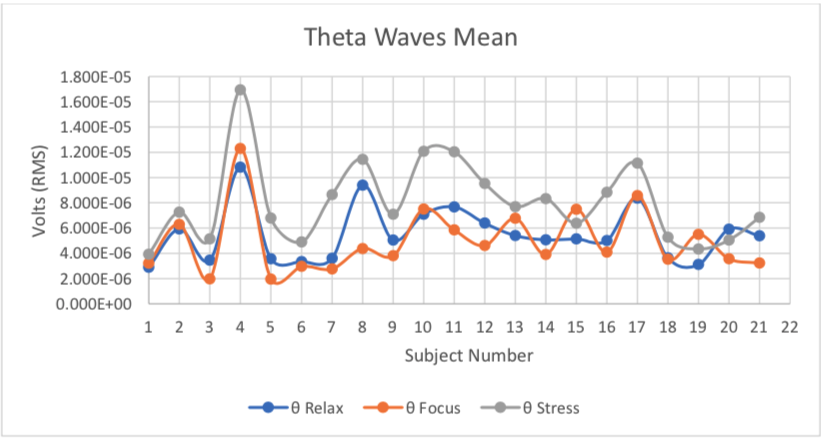}
\caption{Theta wave mean values for each subject.}
\label{ThetaWaveMeans}
\end{figure}

In terms of stress level, the indicators are a high theta wave and low alpha wave ~\cite{seo2010stress,subhani2011eeg}. In figure~\ref{ThetaWaveMeans}, the mean values for theta wave did indeed appear to be greater than in other scenes; that said, it is necessary to realise that the head movement may have affected the data. This may also have been the reason that the alpha wave was not substantially lower, despite a lowered alpha being regarded as an indicator of stress. While this may have been the case, it was clear to see that the theta wave is proportionally higher in some cases, such as in subjects 4 and 10, both of whom had much higher heart rates in the stress scene. This may be an indicator that, despite the spikes in signal due to head movement, the theta wave does appear to be a reliable indicator of stress levels.

\section{Discussion}
While the analysis of the questionnaire and SUS are outside the scope of this paper, some items of note can be mentioned which helped with the interpretation of the sensor data.
Secondary input systems are designed to be as unobtrusive as possible. While the majority of participants reported that the sensors did not have an effect, there were some responses indicating that it was an issue, with one user strongly agreeing that they were conscious of the sensors throughout the experiment. It is also interesting to note that while a number of users were conscious of the sensors during the experiment, the majority of users didn’t feel that it negatively affected them. This may indicate that there is a level of balance that can be achieved between the number of sensors and the ergonomic effect they have on performance and enjoyment.

Another important point to note is that not all users who agreed or strongly agreed with their reported level of stress showed ECG or EEG signs of this; i.e. not everyone who asserted they were stressed had a much higher heart rate or higher theta wave. This may indicate that a person’s own perception of relaxation and stress does not always correlate to their physiological signals; other secondary input sensor types (e.g. GSR/EDA) may have shown a different result, but at the very least is poses an interesting question and merits further study. 

While there were some limitations with the experiment, namely head movement causing spikes in EEG data, the use of a single dry-electrode EEG, and the ECG requirement to keep both hands on the contact sensors – the results are promising, and indicate that heart rate, alpha waves, and theta waves may be viable sources of secondary input.  

These merit further examination in the context of VR education environments, where the measurement of concentration and/or stress may be an important measure of a student's performance in a learning exercise or experience. It would seem particularly suited to task-based or constructivist learning environments, especially those that require the student to operate under uncertainty or stressful situations, examples of which include surgical training, first-responder training, driving/flight training, laboratory tasks, or police/security education.

\section{Conclusion}
A three-scene VR environment was developed to ascertain the suitability of incorporating EEG and ECG biosignals as sources of secondary input data. The three scenes covered relaxation, concentration/focus, and stress, each of which result in different physiological responses from users. This experiment has shown that the measurement of heart rate, alpha waves and theta waves as captured by the sensors are good indicators of a user's emotional response to stimuli and the virtual environment, and their level of engagement in the system.
\bibliography{ms}
\end{document}